\documentclass[12pt,preprint]{aastex}
\def\Msun{$M_\odot$}
\def\msun{$M_\odot$}
\def\Mc{$M_c$}
\def\C12{^{12}C}
\def\N14{^{14}N}
\def\O16{^{16}O}
\def\Mg24{$^{24}$Mg}
\def\Ne20{$^{20}$Ne}
\def\F20{$^{20}$F}
\def\Na24{$^{24}$Na}
\def\ocen{$\omega$~Cen}
\def\fcno{F$_{CNO}$}
\def\simgt{\lower.5ex\hbox{$\; \buildrel > \over \sim \;$}} 
\begin{document} 

\title{Self--enrichment in Globular Clusters: is there a role for the super--asymptotic
giant branch stars?}

\author{M.~L. Pumo\altaffilmark{1,2,3}, F. D'Antona\altaffilmark{1},  P. Ventura\altaffilmark{1}}

%\offprints{} \mail{}
\affil{\altaffilmark{1}INAF - Osservatorio Astronomico  di Roma, via Frascati
33, 00127 Roma, Italy; mlpumo@astrct.oact.inaf.it, dantona@mporzio.astro.it, ventura@mporzio.astro.it}
\affil{\altaffilmark{2}Dip.S.F.A. - Universit\`a di Palermo, via Archirafi 36, 90123 Palermo, Italy}
\affil{\altaffilmark{3}INAF - Osservatorio Astrofisico di Catania, via S. Sofia 78, 95123 Catania, Italy}
\date{Received ... ; accepted ...}

\begin{abstract}
In four globular clusters (GCs) a non negligible fraction of stars can be interpreted only as a very helium rich population. The evidence comes from the presence of a ``blue" main sequence in \ocen\ and NGC~2808, and from the the very peculiar horizontal branch morphology in NGC~6441 and NGC~6388.
Although a general consensus is emerging on the fact that self--enrichment is a common feature among GCs, the helium content required for these stars is Y$\simgt$0.35, and it is difficult to understand how it can be produced without any ---or, for \ocen, without a considerable---associated metal enhancement.
We examine the possible role of super--AGB stars, and show that they may provide the required high helium. However, the ejecta of the most massive super--AGBs show a global CNO enrichment by a factor of $\simeq$4, due to the dredge--out process occurring at the second dredge up stage. If these clusters show no evidence for this CNO enrichment, we can rule out that at least the most massive super--AGBs evolve into O--Ne white dwarfs and take part in the formation of the second generation stars.
This latter hypothesis may help to explain the high number of neutron stars present in GCs. The most massive super--AGBs would in fact evolve into electron--capture supernovae. Their envelopes would be easily ejected out of the cluster, but the remnant neutron stars remain into the clusters,
thanks to their small supernova natal kicks.
\end{abstract}


\section{Introduction}

The majority of the inhomogeneities in the chemical composition of 
Globular Cluster (GC) stars ---see e.g. \citet{sne99,gratton-araa2004}--- appear due 
to primordial enrichment by hot-CNO cycled material processed in
stars belonging to a first stellar generation. Either massive AGB envelopes
subject to hot bottom burning, \citep*[e.g.][]{ventura2001, ventura2002}
or the envelopes of massive fastly rotating stars (FRMS) \citep{decressin2007} are an ideal 
environment to manufacture elements through nuclear reactions in which proton 
captures are involved. A second stellar generation would then be born from the
ejecta of these stars, either directly, or diluted with pristine material.
There are however problems with these scenario. For the AGB progenitors, problems
concern the chemistry of the anomalous stars (see \citealt{denis2003,ventura2004,
fenner2004}, but consider also \citealt{venturadantona2005a}), and the requirements for the
mass budget necessary to produce a second stellar generation that, today, is generally
about as numerous as the first one \citep{dantona-caloi2004, bekki-norris2006, 
prantzos2006}. 
The FRMS model is somewhat less problematic concerning the constraints on the IMF, 
but poses, in any case, severe problems for ``normal'' GCs, that show a unique
metallicity for all stars (with the notable exception of \ocen): this model requires 
that the envelopes of these stars contribute to the second star formation stage, 
but the supernova ejecta are expelled from the clusters and do not enrich the second
generation matter by heavy elements. 

Both scenarios predict that the
stars showing chemical anomalies also must have {\it enhanced helium abundance}. 
Helium and the hot--CNO products in the FRMS come to the stellar surface by means of the chemical
mixing associated with the transport of angular momentum through the stellar radiative
layers, during the phase of hydrogen burning. 
In the case of AGBs, the helium enrichment is mainly due to the
second dredge up (2DU) which precedes the thermal pulse phase. Also the third dredge
up (3DU) contributes to helium enrichment in the envelopes, but a further observational constraint
is that the sum of CNO abundances remains remarkably constant in normal and anomalous
stars \citep[e.g.][]{cohenmelendez2005}. Consequently, the masses which
can be responsible for the chemical enrichment must be high enough to be subject
to a few episodes of the third dredge up, that brings into the envelope the products of
helium burning. 

A variation in helium content is immediately reflected on the
morphology of the horizontal branch (HB), which amplifies any evolutionary difference
among the cluster stars. Helium enrichment by a small factor (up to Y$\sim 0.28 - 0.30$)
allowed in fact an easy interpretation of
some puzzles posed by the HB (blue tails, gaps) \citep{dantona2002,dantona-caloi2004}. 
Helium variations from the MS location are much harder to be detected,
but the clustering of $\simeq$20\% of \ocen\ stars into 
a ``blue'' MS \citep{bedin2004} was soon interpreted 
as a MS having an abnormally 
huge helium content (Y$\sim$0.40) \citep{norris2004, piotto2005}. 
In contrast to \ocen, NGC~2808 does not show a metallicity spread 
\citep{carretta2006}. Nevertheless, evidence for the presence of a blue MS in
this cluster has been found by \cite{dantona2005}, who interpreted
it, again, as built up by stars having Y$\sim$0.40, and were able
to correlate them with the bluest side of the HB. \cite{piotto2007} showed
that the MS of this cluster is actually made up by {\it three} separate MSs, the bluest
of which, including 10--15\% of stars, must have a helium content Y$\sim 0.35 - 0.40$.
The presence of blue MS has not yet been found in other clusters: observationally
it is a very difficult problem, and, in addition, this may be a characteristic
of the most massive clusters only. 

Huge helium overabundances are also needed to explain all the
main features of the anomalous HB of the metal rich clusters NGC~6441 and NGC~6388
\citep{caloi-dantona2007a} including: 1) the long extension in luminosity of the red clump;
2) the fact that RR Lyrs have a very long average period, which
is unusual for a cluster of high metallicity;
3) the extension into the blue of the HB.
This analysis also requires that $\sim 15$\% of the HB population has Y$>$0.35.

All these observations and their interpretation in terms of helium enrichment must be taken very
seriously, but no chemical enrichment mechanism is able to produce the huge
amount of helium required for about 15\%--20\% of the stars in these four clusters
without dramatically impacting upon the metal
abundance \citep{karakas2006,romano2007}. 
It is clear that the presence of this extreme population 
requires some very particular dynamical and chemical conditions to be understood.
Nevertheless, the first step is to look for progenitors which may at least provide
the required high helium abundance in their ejecta.

The most massive FRMS may be able to eject huge quantities of helium 
\citep{decressin2007}. 
Concerning the AGB stars, \cite{karakas2006} pointed out that Y$\sim$0.40 is not predicted
by AGB models, and this could, in the end, rule out AGB stars as progenitors of the
second stellar generation. A scarne Y$\sim$0.29 was the maximum abundance computed from AGB 
evolution in the models by \citet{ventura2002} for 5.5\Msun, while a single value 
Y=0.375 is quoted by \cite{karakas2006} for a 5\Msun\ model by \cite{campbell2004}. 
P. Ventura \& F. D'Antona (in preparation) have now completed new computation of massive AGBs, 
whose results revise upward the previous ones (see later). 
The maximum AGB mass in these recent computation is 6.3\Msun, and 
the possible role of super--AGB stars has not yet been considered.
Such models are difficult to be computed, as Carbon burning in semi--degeneracy
has to be followed, and the uncertainty in the mass loss rate 
hampers any conclusions on their final evolution, either towards electron--capture 
supernova (ecSN) or towards O--Ne white dwarfs. 

In the light of recent super--AGB stars models described in \cite{pumo2006}, 
in this paper we discuss the possible role of these stars for the self--enrichment of GCs.
We discuss both the helium abundance and the total CNO abundance versus core mass
which can be built up by the most recent AGB and super--AGB models. 
We find out that the matter lost by the envelopes of super--AGBs may indeed be a
good candidate for the high helium population of the four GCs discussed above.
However, in the {\it higher mass tail of the stars developing a
O--Ne core} the total CNO is increased by a factor $\sim 4$. Observations of the
super He--rich stars may be able to check this result.

If all the stars in the GCs having a high--He population have standard CNO, as found 
for all the chemically anomalous stars in many GCs, the most massive 
super--AGBs must evolve into ecSNe. This conclusion would be 
of further support to the idea that the 
presence of large population of neutron stars in GCs implies
that they were born with supernova--kicks low enough not to be ejected by the cluster, as
expected by the ecSN event \citep[e.g.][]{ivanova2007}.

\section{Super-AGB stellar evolution}
AGB stars develop a degenerate carbon--oxygen core and mass loss prevents them
to reach the mass limit for degenerate explosive carbon ignition. Super--AGBs are defined
as stars which ignite carbon in conditions of semi--degeneracy, thus non explosively,
but are not able to ignite hydrostatic Neon burning 
in the resulting O--Ne core. Consequently, degeneracy increases
in the core and these stars may
undergo thermal pulses \citep[e.g.][]{iben1997,ritossa1999, siess2006} and lose mass as
``normal'' (but quite massive and luminous) AGB stars. 
Depending on the competition between the 
mass loss rate and the core growth, they may then evolve into massive O--Ne white 
dwarfs \citep{nomoto1984} or as ecSN, 
electrons being captured first by \Mg24\ and \Na24\ then by \Ne20\ and \F20\
nuclei when the core mass reaches the Chandrasekhar mass.
The 2DU occurs also in super--AGBs and, as in the massive AGBs, it reduces
the mass of the H--exhausted core \citep[e.g.][]{pumo-siess2006}, while the helium abundance increases in the external
hydrogen envelope \citep[e.g.][]{siess-pumo2006}.
If mass loss is strong enough to avoid the super--AGB evolving to ecSN, as soon as
super--AGBs evolve in the GCs, the supernova epoch
is finished, and we may hypothesize that the second phase of star formation begins
from the super--AGB ejecta. 

\section{Helium and CNO in massive AGBs and super--AGBs}
\begin{figure}
\epsscale{.95}
\plotone{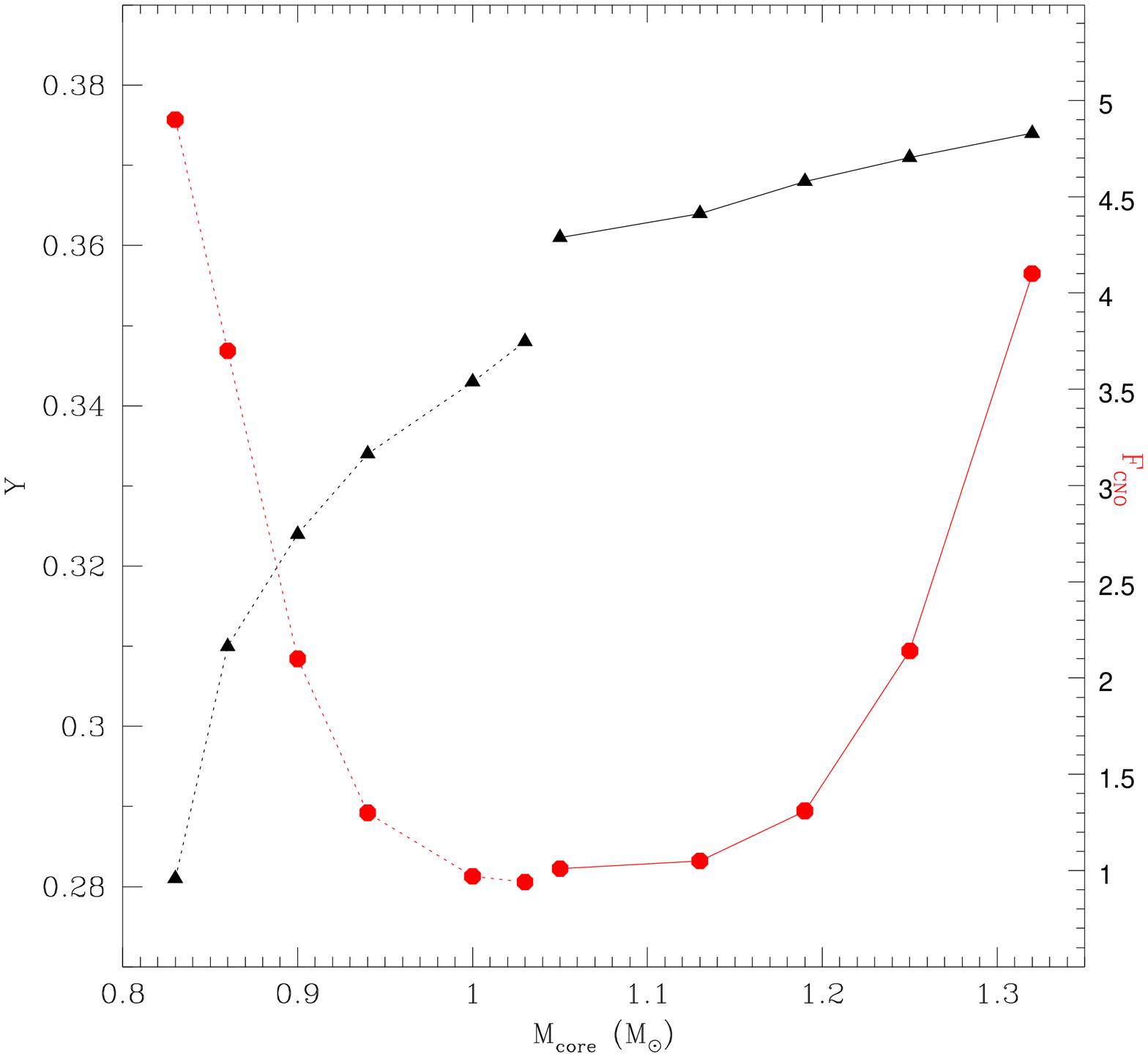}
\caption{The triangles plus dashed line show the helium content in the ejecta 
of AGB stars as function
of the C--O core mass, in the computations by P. Ventura \& F. D'Antona (in preparation).
The initial masses are, from left to right, 4, 4.5, 5, 5.5, 6, 6.3\Msun. The triangles
plus continuous line show the helium abundance in the envelope after the 2DU in the
super--AGBs by \cite{pumo2006}. Masses from left to right are 7.5, 8, 8.5, 9, 9.5\Msun.
The circles show the ratio of total CNO abundance (\fcno) with respect to the initial value
for the same models.\label{f1}}
\end{figure}

As a function of the Carbon Oxygen core mass \Mc, we report in 
Figure \ref{f1} the helium content in the AGB ejecta of
metallicity Z=$10^{-3}$, for the models from 4 to 6.3\Msun\ by P. Ventura \& F. D'Antona (in preparation)
including $\alpha$--elements enhancement in the opacities, and core
overshooting. An important difference from \citet{ventura2002} models 
is the following: in the present models, also overshooting at the
bottom of the convective envelope is included, adopting the same parameter which
describes core overshooting. This latter is calibrated to account for the 
width of the MS of intermediate mass stars. The effect of the envelope
overshooting is to enhance the extent of the 2DU: fixed the initial mass,
the remnant core mass at the
beginning of the AGB evolution is then {\it smaller}, and the resulting helium content 
in the envelope is {\it larger}, as can be seen by comparing the values of Figure
\ref{f1} with Table 1 in \citet{ventura2002}. 
 
We also plot the helium content at the 2DU for the super--AGB 
models of metallicity Z=$10^{-3}$ from 7.5\Msun\ to 9.5\Msun\
\citep{pumo2006} computed without overshooting.
In this case \Mc\ is the C--O mass after the 2DU. \Mc\ is indeed the important
physical parameter which determines the final fate of the star.
The initial mass difference between the AGB models with overshooting and super--AGB models
without overshooting is, for the same core mass, $\sim$1\Msun. 
For super--AGBs, we do not have the helium mass fraction of the ejecta, but these stars
do not go through efficient episodes of 3DU, so the plotted value is a good approximation.
As we see, the super--AGB models reach a helium content in the ejecta larger
than the standard AGBs, and well approaching the values
needed to be consistent with the super--He rich stars in GCs. The new
AGB models and the super--AGB models shown here, computed by a different
program and input physics \citep{siess2006, siess-pumo2006} 
are in good agreement with each other, at the same core mass.

From Figure \ref{f1} we see that indeed the super--AGBs could be the
progenitors of the very high helium population found in the most massive GCs. 
For a Salpeter's initial mass function, the mass budget in the ejecta from 6.5\Msun
(the minimum mass for super--AGB evolution, considering models with overshooting) 
to 8\Msun (see later) is about 50\% of the mass budget contained
in the ejecta of normal AGBs from 5\Msun to 6.5\Msun. Therefore, from the super--AGBs 
mass budget, a 15\% fraction of very high helium stars can be born, with the same mechanism
which can give origin to another $\sim$30\% of moderately helium rich stars, as in the
cluster NGC~2808 \citep{dantona2005}.
How these stars can form directly from the super--AGB ejecta, so that 
their helium abundance remains as large
as the required Y$\simgt$0.35, is a different problem.

Figure \ref{f1} also shows the total CNO enrichment \fcno, the
sum of CNO abundances with respect to their initial sum, in the ejecta of AGBs
and at the 2DU for the super--AGBs. The AGB enrichment is larger than $\sim$2 for
\Mc$\le$0.9\msun, that is for M$\le$5\msun. As we have noticed, the sum of CNO
abundances apparently is constant (within a factor 2) in GC stars. On this basis, we
should conclude that stars of M$<$5\Msun\ should not contribute to the second stellar
generation --unless their ejecta are diluted with pristine matter, so that the constancy
of CNO is preserved. For the super--AGBs, we also find a sharp increase of \fcno,
reaching a factor 4 for the largest mass in exam. This is due to the occurrence of the
so called dredge--out process in this model \citep{iben1997, ritossa1999}. 
The helium mass fraction and the overall CNO content are the most robust
predictions from these theoretical computations, due to the large uncertainties
associated to both the extension of the 3DU and the strength of
hot bottom burning (HBB) within super--AGB models \citep[e.g.][]{IP06,pumo-siess2006}.
We also comment the individual abundances providing \fcno. The CNO abundances given for $M\leq6.3$\msun
are the average mass fractions of the ejecta. As HBB is very efficient, the values provide
very small carbon ($X_{\C12} \simeq 1.1 \cdot 10^{-4}$ for $4$-$5$\msun models down to
$X_{\C12} \simeq 1.6 \cdot 10^{-5}$ at $6.3$\msun), and low oxygen (from $X_{\O16} \simeq 6.3 \cdot 10^{-4}$ 
at $4$\msun to $X_{\O16} \simeq 10^{-4}$ at $6.3$\msun); nitrogen is very high, 
especially in the $4$\msun, where the effect of the 3DU is more visible ($X_{\N14} \simeq 2.6 \cdot 10^{-3}$).
On the contrary, for the super--AGBs, we consider the values at the 2DU. The carbon abundance
increases from $X_{\C12} \simeq 8.9 \cdot 10^{-5}$ for the $7.5$\msun model to
$X_{\C12} \simeq 8.1 \cdot 10^{-4}$ at $9$\msun. The nitrogen is about constant 
($X_{\N14} \simeq 2.4 \cdot 10^{-4}$), whereas
the oxygen abundance is constant ($X_{\O16} \simeq 3.8 \cdot 10^{-4}$) up to
$8.5$\msun, and increases up to $X_{\O16} \simeq 4.4 \cdot 10^{-4}$ for the 
$9$\msun model: this is due to the deeper sinking of the 2DU when increasing the mass. In the $9.5$\msun model, the dredge--out process
rises $X_{\C12}$ to $\simeq 1.8 \cdot 10^{-3}$ and $X_{\O16}$ to $\simeq 7.7 \cdot 10^{-4}$. The following evolution
will reduce oxygen and carbon and increase nitrogen, if HBB is efficient in these stars as it is in the massive AGBs 
(this depends on the convective model adopted; see \citealt{venturadantona2005a}). 
As for the 3DU, there are different and 
incomplete literature results \citep[see][for summary]{IP06}. In any case, it can only act to increase the values of 
total CNO abundances given in Figure \ref{f1}. 

How many stars would be affected if all the super--AGBs contribute to the second
stellar generation? If all the super--AGBs contribute to the very helium rich
population, the fraction of high--CNO stars, assuming a Salpeter's IMF, would
be $\simeq$10\%! This is a small fraction of the cluster stars, if we consider 
that the super--He rich stars are 15--20\% of the total population, but 
spectroscopic observations of a large sample of stars could falsify this hypothesis.

\section{Do the most massive super--AGBs explode?}
Should the constancy of CNO be observationally confirmed also for the most helium rich
population showing up in the quoted 4 clusters, we will be able to 
conclude that the most massive super--AGBs did not
take part in the process of forming the second stellar generation in GCs.
This may mean: 1) that there were reasons, independent from the super--AGB evolution,
by which the self--enrichment process was effective either only at earlier ages (due to
the envelopes of massive stars) or at a later age (during the normal, massive AGB evolution)
2) or that the self--enrichment was forbidden by the injection in the cluster gas of the
energy due to the ecSN explosion of the stars which undergo the dredge out.
Although the explosion energy of such event is significantly lower than inferred for
core collapse SN \citep{dessart2006}, it is probably more than sufficient to expel the 
SN ejecta from most clusters.
The conclusion then would be that at least a fraction of super--AGBs
must evolve into ecSN, reinforcing the idea that ecSN channel for 
super--AGB stars could occur \citep[see e.g.][]{siess-pumo2006,pumo-siess2006,Poelarends2007}.

\section{Neutron stars in Globular Clusters}
The final fate as ecSN of ---at least a fraction--- of super--AGBs
is also required by the recent suggestion that NS in GCs 
should be mainly formed by the ecSN channel.

The high number of neutron stars (NS) in GCs, whose presence is 
revealed mainly by the very high
number of millisecond pulsars, remains a puzzle. The massive
GC 47Tuc was predicted to contain more than 1000 NS \citep[see e.g.][]{pfahl2002}. 
Now this figure can perhaps be reduced to $\sim$300--600 
\citep{heinke2005}. The velocity distribution of young pulsars in the Galaxy shows that,
at formation, the NS receives a ``natal kick'', most likely due to the asymmetry in the
supernova ejecta \cite{fryer2004}. \cite{pfahl2002} have shown that only up to 8\% of NS
can be retained in the clusters. The kick velocity distribution which they used \citep{hp1997}
has been recently updated by \cite{hobbs2005}, who find a higher mean velocity and velocity dispersion. 
Using these new data, simulations of the retention factor
by \cite{ivanova2007} show that almost no NS can be retained for an escape velocity
of 40Km$\cdot$s$^{-1}$. A natural way to solve this problem, 
is to hypothesize that there is
a stellar population which evolves into ecSN, that probably have {\it much smaller} natal
kicks \citep{ivanova2007}, possibly proportional to their reduced explosion energy.
This is also suggested by the existence of a class of massive 
X--ray binaries with small eccentricities.
\cite{vdh2007} also supports the evidence of two channels for NS formation ---with 
either small or large natal kicks--- from an exam of the double neutron star binaries.

\acknowledgments

We are grateful to R. G. Izzard for his helpful discussions.

\clearpage

\end{document}